\documentclass{ecai} 

\usepackage{latexsym}
\usepackage{amssymb}
\usepackage{amsmath}
\usepackage{amsthm}
\usepackage{booktabs}
\usepackage{enumitem}
\usepackage{graphicx}
\usepackage{color}

\usepackage{tabularx}

\usepackage[export]{adjustbox}
\usepackage{multirow}
\usepackage{rotating}
\usepackage{float}
\usepackage{subfig} 
\usepackage{transparent}
\usepackage{placeins}
\usepackage[linesnumbered,ruled]{algorithm2e}
\newcolumntype{L}[1]{>{\raggedright\let\newline\\\arraybackslash\hspace{0pt}}m{#1}}
\newcolumntype{C}[1]{>{\centering\let\newline\\\arraybackslash\hspace{0pt}}m{#1}}

\newcommand{\BibTeX}{B\kern-.05em{\sc i\kern-.025em b}\kern-.08em\TeX}

\begin{document}


\begin{frontmatter}


\paperid{10} 


\title{Exploiting Risk-Aversion and Size-dependent fees \\ in FX Trading with Fitted Natural Actor-Critic}


\author[A]{\fnms{Vito Alessandro}~\snm{Monaco}}
\author[B]{\fnms{Antonio}~\snm{Riva}}
\author[B]{\fnms{Luca}~\snm{Sabbioni}\thanks{Corresponding Author. Email: luca.sabbioni@mlcube.com.}} 
\author[B]{\fnms{Lorenzo}~\snm{Bisi}}
\author[C]{\fnms{Edoardo}~\snm{Vittori}}
\author[C]{\fnms{Marco}~\snm{Pinciroli}}
\author[C]{\fnms{Michele}~\snm{Trapletti}}
\author[B,  D]{\fnms{Marcello}~\snm{Restelli}}

\address[A]{Tinaba, Milan, Italy}
\address[B]{ML cube, Milan, Italy}
\address[C]{Intesa Sanpaolo, Milan, Italy}
\address[D]{Politecnico di Milano, Milan, Italy}


\begin{abstract}
In recent years, the popularity of artificial intelligence has surged due to its widespread application in various fields. The financial sector has harnessed its advantages for multiple purposes, including the development of automated trading systems designed to interact autonomously with markets to pursue different aims. 
In this work, we focus on the possibility of recognizing and leveraging intraday price patterns in the Foreign Exchange market, known for its extensive liquidity and flexibility. Our approach involves the implementation of a Reinforcement Learning algorithm called Fitted Natural Actor-Critic. This algorithm allows the training of an agent capable of effectively trading by means of continuous actions, which enable the possibility of executing orders with variable trading sizes. This feature is instrumental to realistically model transaction costs, as they typically depend on the order size. Furthermore, it facilitates the integration of risk-averse approaches to induce the agent to adopt more conservative behavior. The proposed approaches have been empirically validated on EUR-USD historical data.
\end{abstract}
    
\end{frontmatter}


\section{Introduction}
\label{sec:introduction}
The \textit{Foreign Exchange market} (Forex or FX) represents the largest and most liquid market in the world. Because of its high trading volumes, which ensure low transaction costs, it attracts a growing number of potential investors every day. 

Due to the improvement of computational resources and the increasing popolarity of Machine Learning (ML) techniques, the majority of these market players have started exploiting Artificial Intelligence (AI) algorithms to develop profitable automatic trading systems \cite{huang2019automated,lucarelli2019deep,almahdi2017adaptive}. A trading task can be naturally described as a sequential decision problem: market actors iteratively interact with the market by means of trades, observing 
the market state, and receiving a profit or a loss as feedback.
The presence of transaction costs that make necessary to plan on a multi-step horizon, and, thus, to formulate the problem as a as a Markov Decision Process (MDP, \cite{puterman2014markov}).
Since the financial market is hard to model, this problem is more effectively tackled by implementing Reinforcement Learning (RL) algorithms \cite{sutton2018reinforcement}. 

The application of RL techniques to Forex trading has proven successful since the seminal work of \citep{moody2001learning}. In recent times, value-based algorithms employing tree-based function approximators have  demonstrated particularly encouraging outcomes \cite{bisi2020foreign, riva2021learning, riva2022addressing}. These methods have the advantage to be particularly easy to train, since they do not suffer from instability issues, which are the curse of complex neural networks. Therefore, practitioners employing such methods do not have to worry about convergence issues and need only to tune a limited number of hyper-parameters. However, a noteworthy drawback of value-based algorithms lies in their reliance on discrete action space definitions. 

Indeed, these methods provides an estimate of the optimal action-value function by iteratively applying an empirical version  of the Bellman optimal operator. This procedure requires to compute the maximum among the action values at each update. When the action space is continuous, though, this operation must be performed through the discretization of the action space, which requires large computational costs. For this reason, practitioners usually prefer to rely on policy-based methods \cite{schulman2015trust, schulman2017proximal, haarnoja2018soft} for continuous actions tasks.

Consequently, since trading is one of those tasks where considering continuous action spaces enables a more realistic representation of the environment and expands the range of viable solutions, implementing value-based methods encounters significant limitations despite yielding promising results. For instance, the agent is typically constrained to execute fixed-size trading orders. This model simplification impedes the inclusion of the realistic assumption that transaction costs are contingent upon the order size.
Indeed, the larger the trade size (in the order of million dollars), the more difficult it becomes to find enough immediate liquidity for the requested volume, resulting in higher transaction costs. Therefore, the common hypothesis of fixed spreads and costs can hold only when considering limited maximum allocations.

Moreover, continuous actions would offer the opportunity to seamlessly integrate risk-averse approaches, which hold particular significance among financial traders. Their interest lies not only in maximizing returns but also in effectively managing and mitigating investment risks.

Despite the potential benefits of policy-based methods, the learning process of complex parametric policies may be harder to tune with respect to the fitting of non-parametric estimators, which usually requires only to select a few hyper-parameters.  

Based on these considerations, we decided to adopt an \textit{Actor-Critic} (AC) architecture that synergistically combines value-based and policy-based elements to leverage their respective strengths and enhance overall performance. More precisely, actor-critic models employ a parametric policy (the \textit{actor}) for action selection and a value approximator (the \textit{critic}) for value function estimation. In the context of our specific application to Forex trading, this hybrid approach enables us to inherit the promising performance of non-parametric value-based algorithms while retaining the ability of policy-based methods to generalize to continuous actions.

\paragraph{\textbf{Contributions}}
The primary contribution of this paper lies in the innovative application of the \textit{Fitted Natural Actor-Critic} (FNAC, \cite{melo2008fitted}) algorithm to a Forex trading environment. The adoption of an actor-critic approach proves instrumental in effectively learning effective policies for a variety of trading tasks that require a continuous action space.
It is paramount to notice that, differently from more common actor-critic approaches \cite{schulman2015trust, schulman2017proximal, haarnoja2018soft} requiring neural network critics for their guarantees to hold, FNAC allows to select any kind of function approximator as a value function estimator. Therefore, it permits to exploit the advantages of tree-based approaches also in the continuous actions setting.
This kind of action space allows to adopt a more realistic setting for transaction costs and to incorporate risk-aversion techniques to induce more effective behaviors in the models. Consequently, our approach not only demonstrates the practicality of FNAC but also enables a more sophisticated representation of Forex trading dynamics, enhancing the overall robustness and applicability.

\paragraph{\textbf{Outline}}
This paper is organized as follows: in Section~\ref{sec:background} we introduce the Reinforcement Learning background focusing on the description of FNAC algorithm. In Section~\ref{sec:related} we survey literature related to our problem, which is illustrated in detail in Section~\ref{sec:problem}. In the same section we present the algorithm implementation details for the risk-neutral setting and the risk-averse one.
The experimental results are presented in Section~\ref{sec:results}, while some final considerations are reported in Section~\ref{sec:conclusions}.
\section{Background}
\label{sec:background}
Reinforcement Learning is the branch of ML that focuses on training agents to learn the optimal behavior to achieve a specific goal while interacting with an external environment. This interaction consists of a sequence of decisions or actions, and it is typically modeled through the MDP framework. An MDP is a stochastic dynamical process defined by the tuple $\langle \mathcal{S}, \mathcal{A}, P, R, \gamma, \mu \rangle $, where $\mathcal{S}$ is the set of states, $\mathcal{A}$ is the set of actions, $P$ is the state-transition probability function defined as $P(s'|s, a)$, R is a reward function, $\gamma \in [0,1]$ is a discount factor for future rewards, and finally $\mu$ is the distribution of the initial state.

The agent selects its actions based on a \textit{policy} $\pi$ that assigns each state a probability distribution over the action space. 
In general, the goal of the agent is to maximize the \textit{return}, that is, the sum of rewards collected through the interaction with the environment. To evaluate a policy $\pi$, we can define the \textit{State-Value} function $V^\pi(s)$ that represents the expected return starting from a particular state $s$ and following $\pi$ thereafter. Similarly, the \textit{Action-Value} function $Q^ \pi(s, a)$ denotes the expected return starting from a particular state $s$, taking an action $a$, and then following the policy $\pi$.

Therefore, solving an RL task involves determining the optimal policy, that is, the policy that maximizes the expected return. When both the transition model $P$ and the reward function $R$ are known, the MDP can be solved by employing the dynamic programming approach. It decomposes the original problem into two tasks - \textit{policy evaluation} and \textit{policy improvement} - and finds the optimal policy by iteratively addressing them.

However, the complete knowledge of the MDP is not available in most real-world applications. Moreover, it is common to deal with extremely large or even continuous state-action spaces. Therefore, in these cases, one has to adopt RL techniques that exploit function approximation methods to learn how to generalize the experience gathered from the observed agent-environment interaction. Two are the main approaches reported in the literature: policy-based algorithms and value-based algorithms. The former directly build a parametrization of the policy function and optimize it to maximize the cumulative reward. Specifically, the policy approximator exploits the \textit{Policy Gradient theorem} \cite{sutton1999policy} to update the parameters. On the other hand, value-based algorithms approximate the value functions and derive the optimal policy indirectly by selecting the actions associated with the highest value estimates.

Both approaches have strengths and weaknesses that make them suitable for different contexts. For instance, one of the advantages of policy-based methods is their ability to handle continuous action spaces seamlessly, making them well-suited for tasks where discretizing the action space is not straightforward or even disadvantageous. However, they are inherently sample inefficient, often requiring more data to converge to the optimal policy, and prone to getting trapped in local optima, especially in high-dimensional and complex environments. On the contrary, value-based methods are well-known for their sample efficiency and ease of implementation. Nevertheless, challenges arise when handling continuous action spaces, which require discretization or approximation techniques. Additionally, large state spaces can pose computational challenges in accurately estimating values for each state-action pair.

Actor-critic methods represent a hybrid approach that combines the strengths of policy-based and value-based paradigms to overcome some of their limitations. Specifically, they employ a first approximator (the \textit{actor}) to model the policy and select actions and a second one (the \textit{critic}) to estimate the value function.

\subsection{Fitted Natural Actor-Critic}

\label{subsection:fitted_natural_actor_critic}
Fitted Natural Actor-Critic is an extension of the Natural Actor-Critic algorithm \cite{peters2008natural}, one of the SotAs in the framework of actor-critic methods. Specifically, it modifies the TD-based critic to enable the use of general value function approximators, employing the natural gradient to update the policy of the actor. Unlike vanilla gradients, natural gradients usually ensures a faster convergence by following the steepest ascent direction in \textit{Riemannian space} \cite{ollivier2017information}:
\begin{equation}
\label{eq:riemannian_space}
    \widetilde{\nabla}_\theta \textit{f}(\theta) = G(\theta)^{-1}\nabla_\theta \textit{f}(\theta),
\end{equation}
where $G(\theta)$ is a positive definite matrix named \textit{metric tensor}. A common choice for this matrix in machine learning, and specifically in the natural gradient framework, is the \textit{Fisher Information Matrix} $F(\theta)$ given by:
\begin{equation}
\label{eq:fisher_information_matrix}
    F(\theta) = \mathbb{E}_{\tau \sim p_\theta}[\nabla_{\theta}\log p_\theta(\tau) \nabla_{\theta}\log p_\theta(\tau)^T],
\end{equation}
where $p_\theta$ is a probability distribution conditioned by the value $\theta$.
Using a compatible function approximation \cite{sutton1999policy} defined by:
\begin{equation}
\label{eq:compatible_function_approximation}
    q_{w}(s,a) = w^T(\nabla_\theta \log \pi_\theta(s,a)).
\end{equation}
and recalling (\ref{eq:riemannian_space}) and (\ref{eq:fisher_information_matrix}), the natural policy gradient simplifies to:
\begin{equation}
\begin{aligned}
\widetilde{\nabla}_{\theta}J(\theta) &= F(\theta)^{-1}\nabla_\theta J(\theta)=w.
\end{aligned}
\end{equation}
Consequently, we only need to estimate $w$ instead of $G(\theta)$, thus leading to a policy improvement step of:
\begin{equation}
\label{eq:natural_gradient_update_policy}
\theta_{t+1} = \theta_t + \alpha w.
\end{equation}

FNAC uses a dataset $\mathcal{D}$ of samples obtained from the environment: during each iteration, the critic component processes the samples to compute an approximation $\widehat{V}_\theta$ related to the policy $\pi_\theta$. This approximation is then used to estimate an approximated form of the advantage function $\widehat{A}_\theta$ by employing a linear function approximation with compatible basis functions, which is used to perform the natural policy gradient update of the $\theta$ parameters of the actor component.
Specifically, $\widehat{V}_\theta$ is estimated by fitting any type of regressor model and by minimizing the Bellman Error using the samples in $\mathcal{D}$. Furthermore, the corresponding advantage function can be approximated by tackling the following regression problem:
\begin{equation}
\label{eq:nat_gradient_regression}
    w^\star = \arg\min_w \sum_t \frac{1}{\widehat{\mu}(s_t)}(r_t + \gamma \widehat{V}_\theta(s_{t+1}) - \widehat{V}_\theta(s_t) - \phi^T(s_t, a_t)w)^2,
\end{equation}
where $w$ denotes a linear coefficient vector. This vector corresponds to the orthogonal projection of the advantage function $\widehat{A}_\theta$ within the linear space defined by the compatible basis functions $\phi(s_t, a_t)$:
\begin{equation}
\label{eq:compatible_basis_function}
    \phi(s_t, a_t) = \frac{\partial \log(\pi_\theta)}{\partial \theta}(s, a).
\end{equation}
This regression problem can be solved using an ordinary least square approach:
\begin{equation*}
    M = \sum_t \frac{1}{\widehat{\mu}(s_t)}\phi(s_t, a_t)\phi^T(s_t, a_t),
\end{equation*}
\begin{equation*}
    b = \sum_t \frac{\phi(s_t, a_t)}{\widehat{\mu}}(r_t + \gamma \widehat{V}_\theta(s_{t+1}) - \widehat{V}_\theta(s_t)).
\end{equation*}
To finally obtain $w^\star=M^{-1}b$, which is used to update the policy using the natural gradient formula in (\ref{eq:natural_gradient_update_policy}).

\section{Related Works}
\label{sec:related}
The first promising results obtained by applying techniques of RL to Forex trading were demonstrated by \cite{moody2001learning}. The author implemented a \textit{Recurrent Reinforcement Learning} (RRL) algorithm to trade the USD/GBP currency pair. Specifically, the model was trained to maximize a variant of the Sharpe Ratio,  referred to as the \textit{differential Sharpe Ratio}. The RRL agent had the flexibility to take long, neutral, or short positions while being constrained to execute fixed-size orders.

In recent years, instead, value-based methods have shown the most encouraging outcomes. \cite{huang2018financial} and \cite{sornmayura2019robust} successfully applied the Deep Q-Network (DQN) \cite{mnih2013playing} algorithm to exploit price patterns in OHLC market data of several currency pairs. However, DQN's lack of interpretability represents a significant limitation. The Fitted Q-Iteration (FQI, \cite{ernst2005tree}) algorithm overcomes this issue by leveraging decision-tree-based regression methods to enhance explainability. \cite{riva2021learning} further refined FQI by incorporating the concept of action persistence to optimize the control frequency for improved signal-to-noise ratio handling.

In the domain of risk-averse approaches,  \cite{bisi2020foreign} introduced a Multi-Objective formulation of FQI for Forex trading, enabling the agent to adopt more cautious strategies. Moving to policy-based algorithms, numerous examples of risk-averse approaches emerged. \cite{bisi2019risk} incorporated the \textit{mean-volatility} objective function into the TRPO \cite{schulman2015trust} algorithm allowing the agent's learning of risk-averse strategies. The model was trained on S\&P 500 daily price data from the 1980s to 2019. Experimental results highlight how tuning the risk aversion of the agent allows for moving along the Pareto frontier between reward volatility and expected return.
 \cite{yang2020deep} employed instead an ensemble method that combined PPO \cite{schulman2017proximal}, A2C and DDPG \cite{lillicrap2015continuous} algorithms to trade multiple stocks. 
Along with the close prices of each stock, the authors include in the environment state formulation the available balance, the number of shares owned, and other financial indicators. The action space was defined as the interval [-\textit{k}, \textit{k}], where \textit{k} is the number of shares that the agent can buy or sell for each stock.
Finally, they incorporated a transaction cost equal to 0.1\% of the traded value and adopted the financial turbulence index \cite{kritzman2010skulls} as a risk-aversion measure for capturing extreme asset price movements.
Overall, the ensemble strategy involved training and validating agents iteratively to select the best-performing agent for the subsequent trading period.

\begin{figure*}
\centering
\begin{subfloat}
  \centering
  \includegraphics[width=\columnwidth]{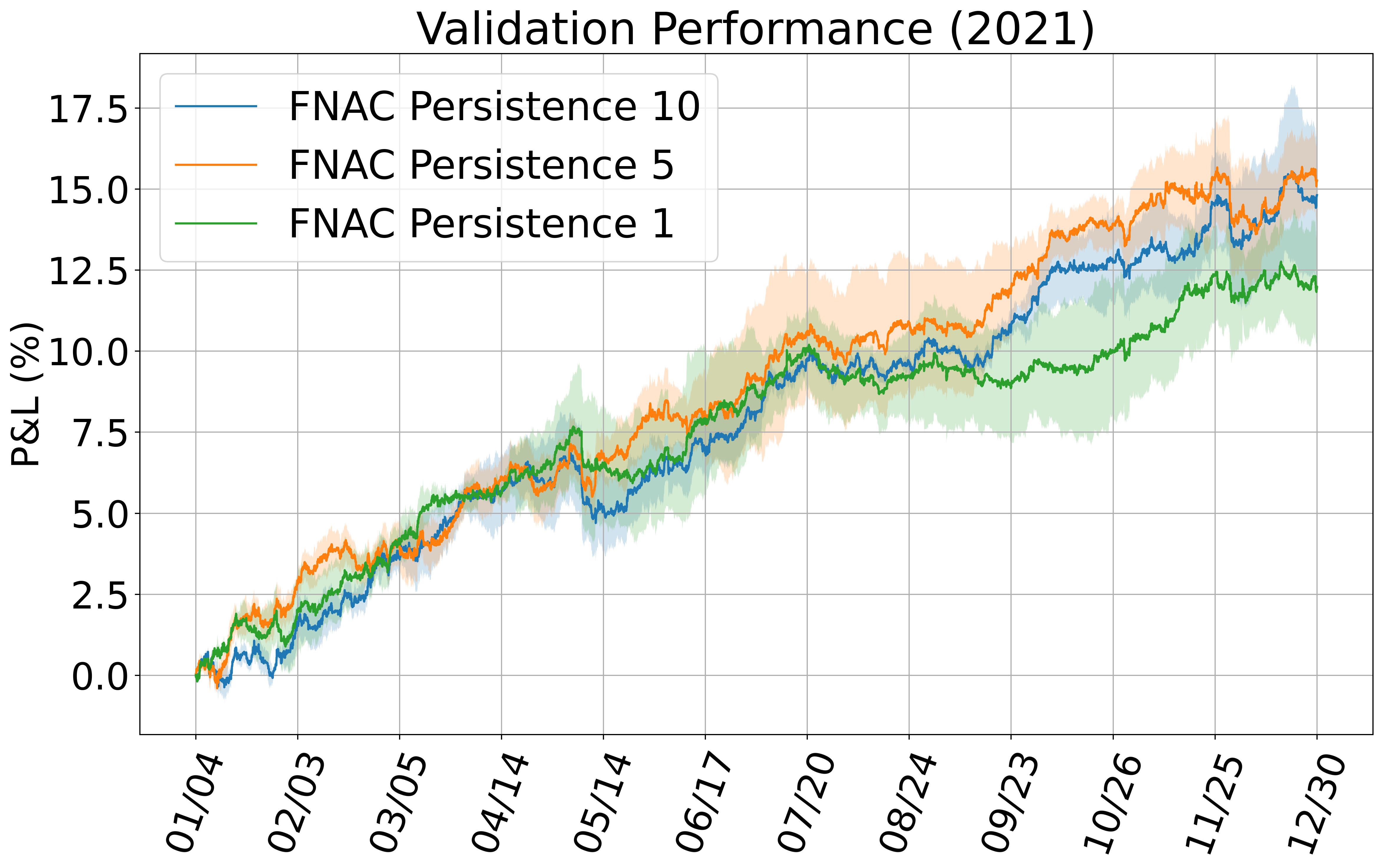}
  \label{fig:fnacdiscrete_2021}
\end{subfloat}%
\begin{subfloat}
  \centering
  \includegraphics[width=\columnwidth]{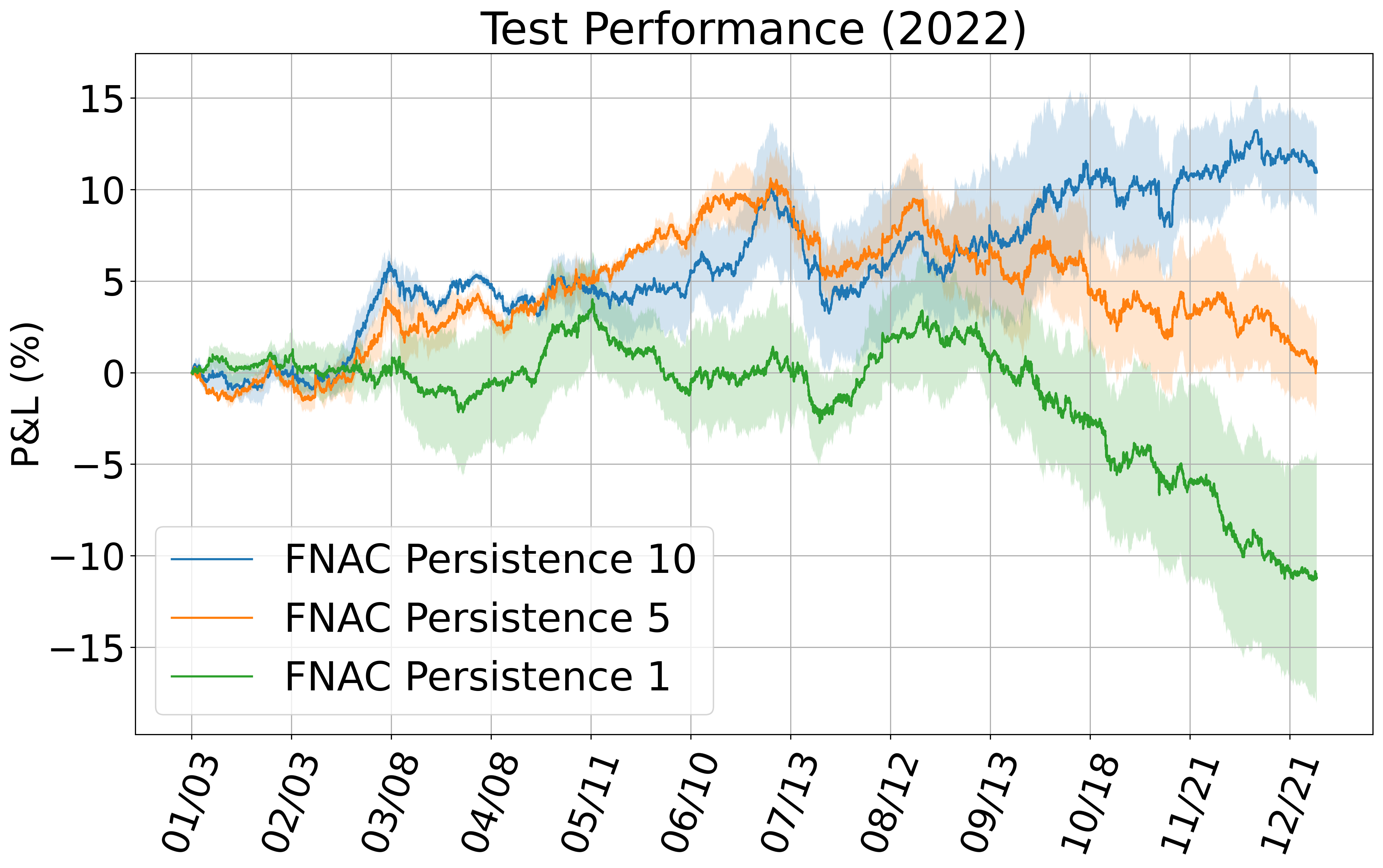}%
  \label{fig:fnacdiscrete_2022}
\end{subfloat}
\vspace*{1mm}
\caption{Expected cumulative returns obtained by the FNAC models trained in the discrete action space trading setting with different action persistence. Plots on the left and on the right show curves, respectively, for  validation and test years. Performance is reported as mean percentages w.r.t. the invested amount (mean $\pm 2 stds$).}
\label{fig:fnacdiscrete_persistence}
\vspace*{6mm}
\end{figure*}
\section{Problem Formulation}\label{sec:problem}
\subsection{Forex Trading Enviroment}
We formalize the Forex trading problem as an episodic task, where an episode corresponds to a trading day consisting of a time window of 600 minutes between 8:00 a.m. and 6:00 p.m. (CET). At the end of each episode, the agent is forced to close its position. There are two reasons behind this choice: first of all, considering fixed-length episodes simplifies the learning problem and enables working in the undiscounted setting (i.e., $\gamma = 1$); secondly, it allows neglecting overnight funding costs.

\subsubsection{Action Formulation}

The action consists of the portfolio allocation the agent wants to keep in the next state.
We allow three possible portfolio allocations: \textit{Short}, \textit{Flat}, and \textit{Long}. Therefore, we defined the allocation as a real number $x \in [-1,1]$, where $x = 0$ denotes the Flat position, while the closer $x$ is to -1 (or 1), the higher the exposure of the Short (or Long) position (up until the maximum exposure which we fixed equal to $100 \text{K}\$$). 
However, in the discrete setting, we assume that the agent can only execute fixed-size orders, thus taking only fully exposed positions. In other words, we define the action space $\mathcal{A}$ as the discrete set $\{-1, 0, 1\}$. On the contrary, in the continuous setting, there are no constraints on the order size, hence the agent can assume any allocations in the range [-1,1], that is,  $\mathcal{A} = [-1,1]$.

\subsubsection{State Formulation}
To effectively model the Forex trading environment as an MDP, it is fundamental to ensure the Markov property holds. This property states that each new state in the process should be independent of its previous states, given the current state and action taken. Since it is computationally unfeasible to include all the information related to past market observations that may affect the transition to the next state, in practice, only a finite window of market information is given as input to the agent. Following \cite{riva2021learning}, we have included in the environment state formulation the last 45 mid-price variations between consecutive minutes, the values of the spread, the day of the week and the minute of the specific trading day. Namely, we will denote the relative variations as $$d1\_n:=\frac{p_{t-n}-p_{t-n-1}}{p_{t-n-1}}, \ n\ge0.$$ Moreover, to take into account the transaction costs we included the current portfolio allocation determined by the previously executed action. 

\subsubsection{Reward Formulation}
The reward received by the agent after performing an action $a_t$ is given by:
\begin{equation}
\label{eq:reward_formulation_persisted}
    r_{t+1} = \underbrace{a_t(p_{t+1} - p_{t})}_{\text{gains/losses}} - \underbrace{c_t|a_t - x_t|}_{\text{trans. costs}},
\end{equation}
with 
\begin{equation}
\label{eq:transaction_cost}
    c_t = g(|a_t-x_t|)\sigma_{t},
\end{equation}
where $p_t$ and $p_{t+1}$ denote respectively the exchange rate at time $t$ and $t+1$, $x_t$ the current portfolio allocation and $\sigma_t$ the current value of the bid-ask spread. Note that, to realistically model the transaction cost as dependent on the order size, we introduced a monotonically increasing function $g: [0,2]\rightarrow \mathcal{Y} \in \mathbb{R}^+$, which maps the order size to a scaling factor of the spread.
This formula strictly generalizes the one in \cite{riva2021learning}, which can be recovered by substituting $g$ with the half of the identity function. 
In summary, the reward is given by the sum of two terms: the first represents the gain (or the loss) due to the exchange rate variation, whereas the second consists of the transaction cost the agent has to pay to change portfolio allocation. Our generalization permits to model a wider range of real scenarios, in which the order size is allowed to change at each time-step.

\subsubsection{Action Persistence}

When a continuous-time problem is transposed into a discrete-time MDP framework, a time discretization and therefore a control frequency are introduced. Choosing accurately the control frequency, especially in the trading framework, is fundamental. Indeed, despite higher frequencies enabling higher control opportunities, hence higher potential profits, on the other hand, the observation noise and the sample complexity might become too high and compromise the learning process. 

To effectively tune the control frequency and optimize the signal-to-noise ratio, we exploit the concept of \textit{action persistence} \cite{metelli2020control}, which consists of repeating each action for a certain number $k\ge1$ of consecutive steps.
Note that, to generalize the reward definition introduced above is sufficient to modify Equation (\ref{eq:reward_formulation_persisted}) by substituting $p_{t+1}$ with $p_{t+k}$. 

\subsection{Algorithm Implementation}
\label{subsection:Algorithm_Implementation}
In this section, we briefly illustrate how we adapted the FNAC algorithm before applying to solve the Forex MDP. We recall from Section \ref{subsection:fitted_natural_actor_critic} that the FNAC architecture consists of two components: the critic, which estimates the value function, and the actor, which estimates the policy. The critic is in turn composed of two layers: the first one is responsible for estimating the value function, while the second one is in charge of approximating the advantage function. For the first layer, we adopted a non-parametric tree-based estimator. These kinds of models have already proven effective in the financial setting \cite{bisi2020foreign, riva2021learning, riva2022addressing}, thanks to their ability of sharply separating the input space, their ease of training, and their explainability.
In particular, we chose to implement an \textit{XGBoost} \cite{chen2016xgboost} regressor, a state-of-the-art tree-based ensemble function approximator, which is also computationally efficient when dealing with large training sets. For the second layer we employed a linear regressor, as prescribed by the FNAC architecture, using ridge regression. Finally, for the actor we adopted a \textit{Feed-Forward Neural Network} (FFNN). It is fundamental to highlight that the actor architecture depends on the action space considered. In the discrete setting, we used an FFNN with one hidden layer and the final output layer composed of three neurons, each corresponding to one of the available actions. Then, using a softmax function, we normalize the output to obtain the probabilities of selecting each action. Instead, in the continuous action space setting, we adopted an FFNN with two hidden layers and a final output layer composed of two neurons corresponding to the mean and the standard deviation of a truncated normal distribution that models the policy and restricts the action sampled to interval $[-1,1]$.


\subsection{Risk-Averse Variants}
To keep the risk of potential losses under control, it is possible to extend the risk-neutral formulation to integrate risk-averse objectives.
In the trading setting this makes sense only when the agent has access to a continuous action space that allows it to hedge risks by means of intermediate levels of allocation.
We considered two risk measures: the Reward Conditional Value-at-Risk (RCVaR, \cite{bonetti2023risk}), which is a reward-based version of the classic Conditional Value-at-Risk (CVaR), and the Mean-Volatility risk measure objective \cite{bisi2019risk}. The latter measures belong to the reward-based risk-measures group, studied in \cite{bisi2019risk}. Optimizing these types of risk-averse objectives allow to reduce the variability of the reward by means of simple transformations of the reward function.
To optimize the RCVaR with level $\alpha$, the transformed reward corresponds to:
\begin{equation}
\label{eq:rcvar_reward_transformation}
    R^{\alpha}_{\pi}(s,a) = \rho - \frac{1}{\alpha} \left(R(s,a) - \rho \right)^{-},
\end{equation}
where $\rho$ is the $\alpha$-quantile of the discounted state-occupancy reward distribution under $\pi$.
To optimize, instead, the Mean-Volatility objective the following reward transformation has to be performed:
\begin{equation}
\label{eq:mean_volatility_reward_transformation}
    R^\lambda_{\pi}(s, a) = R(s, a) - \lambda(R(s, a) - J_\pi)^2,
\end{equation}
where $\lambda$ denotes the parameter that controls the level of risk-aversion, and $J_\pi$ is the normalized expected return.
While the goal of the former risk measure is to maximize the expected return for a tail of pessimistic scenarios, the latter introduces a trade-off between the maximization of the expected return and the minimization of the variance of the observed rewards.
Both these objectives have also been shown to be proxy for the optimization of their more common return-based counterparts.

 \begin{figure}[t] 
    \subfloat[Persistence = 5\label{fig:feat_imp_discrete_5}]{
        \includegraphics[width=0.24\textwidth]{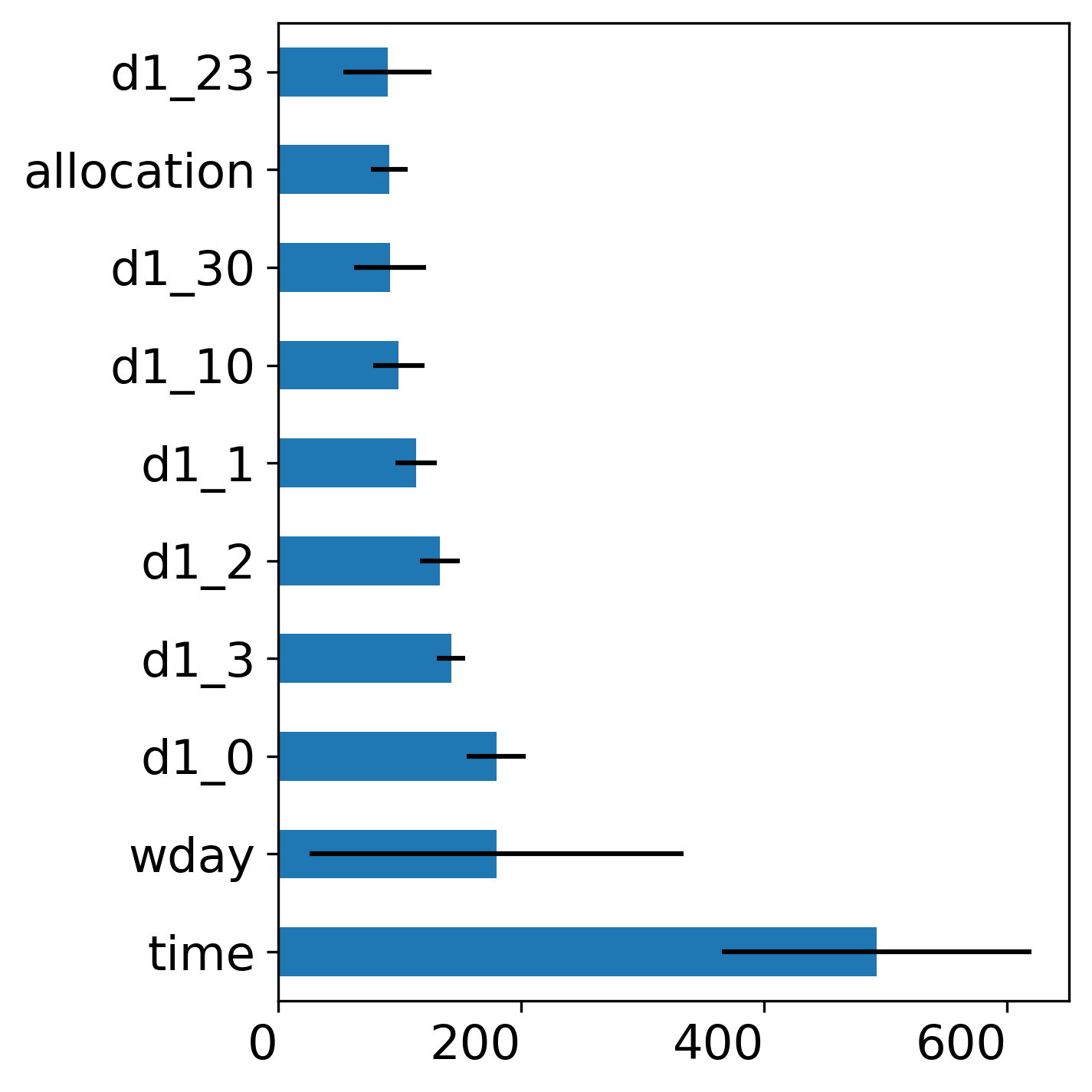}
    }
    \subfloat[Persistence = 10\label{fig:feat_imp_discrete_10}]{
        \includegraphics[width=0.24\textwidth]{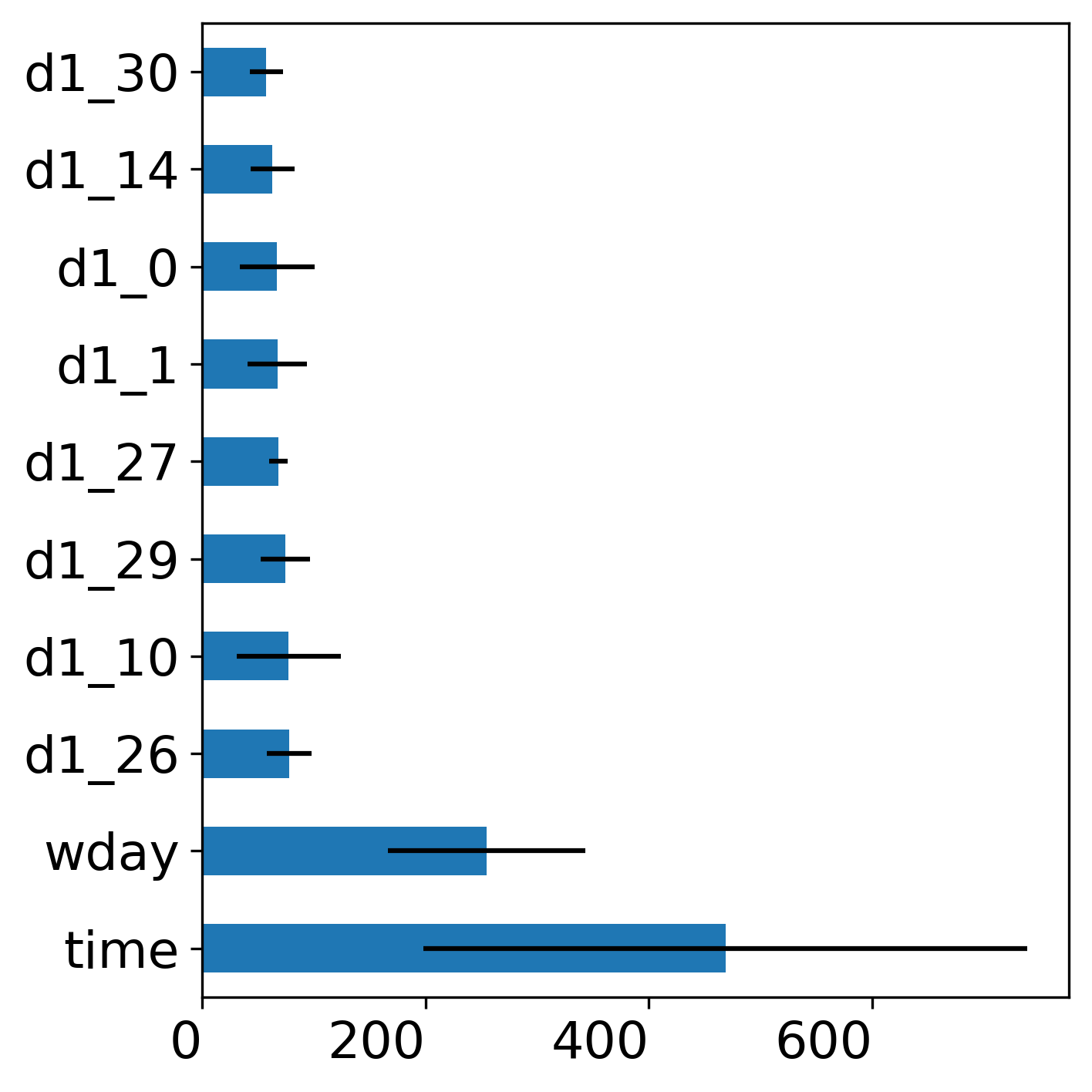}
        }
\vspace*{2mm}
 \caption{Features importance of the FNAC models trained in the discrete action space trading setting with persistence equal to 5 (left) and 10 (right). The importance of a feature is defined as the normalized gain in terms of the Gini impurity index brought by that feature. The mean and the standard deviation of the importance of the ten most relevant features are reported.}
 \label{fig:feat_imp_discrete}
 \vspace*{8mm}
 \end{figure}

\section{Experimental Evaluation}\label{sec:results}
In this section, we describe the model validation procedure adopted to select the best set of FNAC hyperparameters to then illustrate the performance achieved by testing the learned trading strategies on out-of-sample EUR/USD market data in different scenarios. We start presenting the results obtained in the \textit{discrete} action space setting, which were instrumental in determining which performance level FNAC could reach in this simpler framework.
Subsequently, we move to the \textit{continuous} action space setting and analyze the performance of the agent in three trading scenarios: (1) the risk-neutral setting with \textit{fixed} transaction fees, (2) the risk-neutral setting with \textit{variable} transaction fees, and (3) the \textit{risk-averse} setting with fixed transaction fees. It is important to recall that the fixed/variable nature of the transaction fee we refer to regards the dependence on the order size, not the dependence on time. In all the trading scenarios considered, transaction costs vary in time due to the variability of bid-ask spread. 
 
\subsection{Model Selection}
As explained in Section \ref{subsection:Algorithm_Implementation}, we chose an XGBoost regressor for the value function estimation, a Ridge regressor for the advantage function estimation, and an FFNN for the policy approximation. Therefore, to select the best FNAC trading model, we had to tune the most relevant hyperparameters of all the implemented regressors. 

XGBoost is an optimized ensemble method of decision trees that combines gradient boosting and bagging techniques. Thus, two are the sets of hyperparameters that characterize it. The first one includes the parameters that define the construction of the forest of decision trees. Based on our experience and following what is suggested by \cite{riva2021learning}, given a reasonable number of trees that ensures a good trade-off between high computational time and low estimation variance, it is sufficient to tune the $min\,child\,weight$ to regulate the model complexity. The second set of hyperparameters consists of all those parameters related to the gradient boosting framework. Hence, it includes the number of boosting rounds, the learning-rate parameter $\eta$, and the regularization parameters $\gamma$ and $\lambda$. Since all these hyperparameters pursue the same objective in different ways, controlling the model complexity to limit the risk of overfitting, we decided to focus on tuning the learning rate and the number of boosting rounds. 

On the other hand, the Ridge regressor requires validating only the coefficient that controls the L2 regularization. Finally, for the FFNN we tuned the learning rate and the number of hidden parameters.

Given the set of hyperparameters to tune, the validation procedure is straightforward to illustrate. We first split the available data, which includes market observations from 2018 to 2022, into four parts. Then, we use 2018-2019 as the training set, 2020 as the validation set to tune the hyperparameters of XGBoost and the Ridge regressor, 2021 as the validation set for the FFNN, and finally 2022 as the out-of-sample test set.

\subsection{Discrete Action Space Trading Setting}
Before testing our algorithm in the continuous action space trading setting, we investigated the performance of FNAC within the framework of a discrete action space. 
Compared to the continuous action setting, this task is simpler from the training viewpoint. At the same time, it still permits the agent to reach the same optimal policy in a risk-neutral and constant fee scenario. Therefore, it allow us to establish whether passing from the discrete setting to the continuous one is detrimental or not for our algorithm.
We trained FNAC with different persistence values (1, 5, and 10), i.e. acting every 1,5 or 10 minutes, to determine the optimal trading frequency.

In Figure \ref{fig:fnacdiscrete_persistence}, we report the performances obtained by FNAC during 2021 and 2022. Note that, despite the models yielding similar returns at the end of 2021,  the model trained with persistence equal to 10 significantly outperforms the others in 2022. As expected, a lower trading frequency ensured a better signal-to-noise ratio, which increased the generalization capabilities of the model.

One of the strengths of FNAC is its high level of explainability: selecting XGBoost as the regressor to estimate the value function allows us to understand which features contribute the most. For instance, in Figure \ref{fig:feat_imp_discrete}, we report the feature importance of the model trained with persistences equal to 5 and 10. In both cases, the temporal features (\textit{time} and \textit{weekday}) are the most important ones, meaning that our algorithm exploits them to identify recurrent intraday temporal patterns. 

\begin{figure}[t]
    \centering
    \includegraphics[width=0.49\textwidth]{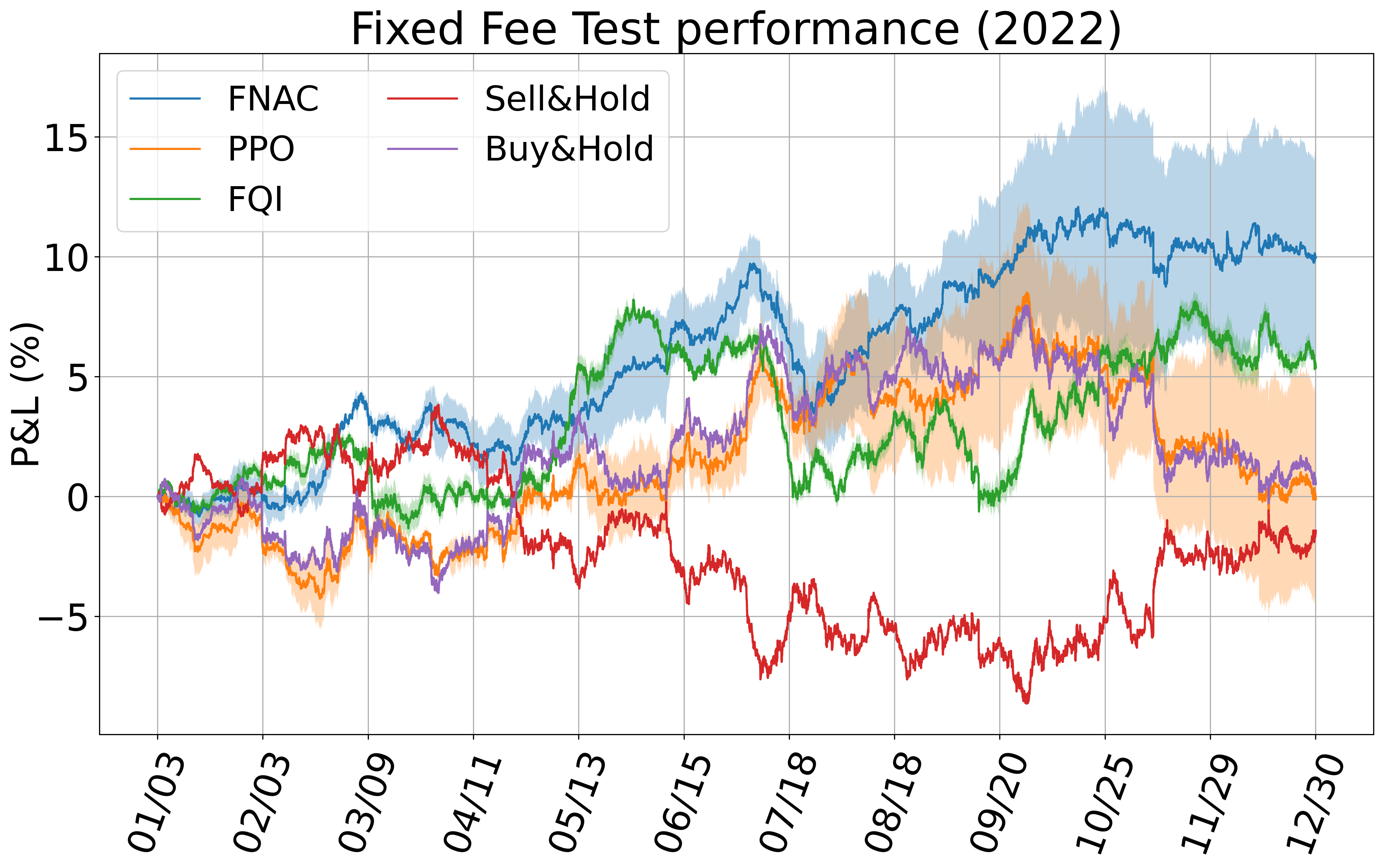}
 \vspace*{1mm}
     \caption{Expected cumulative returns obtained in test by the FNAC model trained in the continuous action space trading setting with fixed transaction fees. The performance of FNAC are compared with those obtained by FQI (trained in the discrete action space trading setting), PPO, and the baseline strategies Buy\&Hold and Sell\&Hold. Performance is reported as mean percentages w.r.t. the invested amount (mean $\pm$ 2 stds).}
    \label{fig:2022continuous_base_plot}
    \vspace*{8mm}
\end{figure}

\begin{figure}[t]
    \centering
    \hspace{0.5em}
    \includegraphics[width=0.48\textwidth]{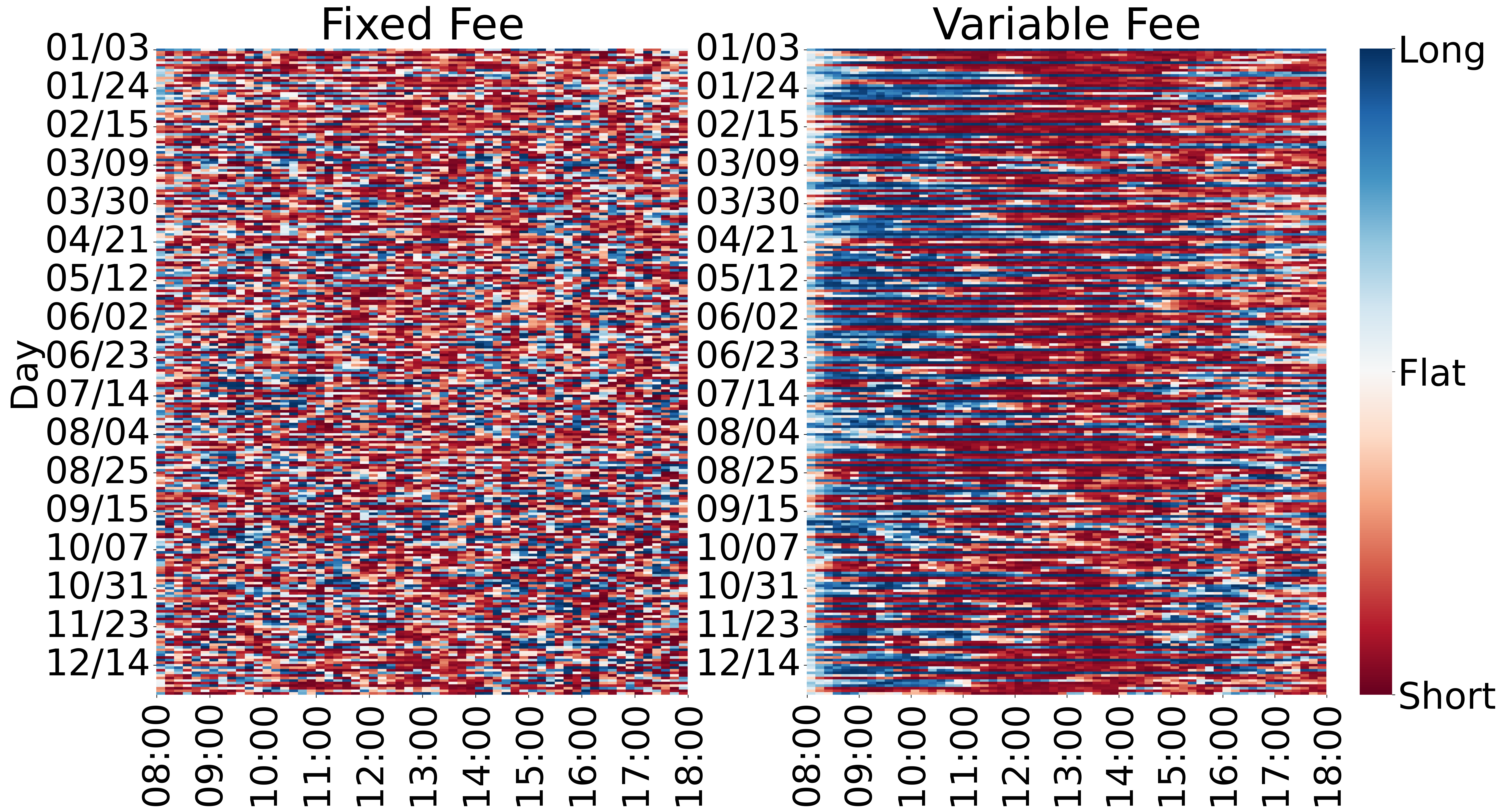}
    \vspace*{1mm}
     \caption{Portfolio allocation selected during the test year by the FNAC models trained in the continuous action space trading setting with fixed transaction fees (left) and variable transaction fees (right). Each row of the heatmaps refers to a different trading day, whereas each column is specific for a trading minute.}
    \label{fig:heatmaps}
    \vspace*{6mm}
\end{figure}

\subsection{Continuous Action Space Trading Setting}
\vspace*{-1mm}
Once FNAC showed promising results in the discrete case, we focused our experiments on the continuous action setting, the real target of our analysis.  Figure \ref{fig:2022continuous_base_plot} shows the results obtained by our method in the fixed fee and risk-neutral scenario, on out-of-sample data. The plot compares the proposed technique with some reinforcement learning baselines (FQI and PPO) and some simple market strategies (Buy\&Hold and Sell\&Hold). Moreover, it also compares the continuous actions version of FNAC with the one employing discrete actions.
The proposed approach obtained slightly worse performances in its continuous form w.r.t. the discrete one. For instance, in the last 4 months of 2022, we noticed higher drawdowns and a significant rise in the standard deviation of the cumulative return. 
Nevertheless, FNAC has proven to achieve significantly better performance compared to all the baselines, included FQI and PPO, for the considered test dataset.

\begin{figure}[t]
    \centering
    \hspace{-1.5em}
    \includegraphics[width=0.49\textwidth]{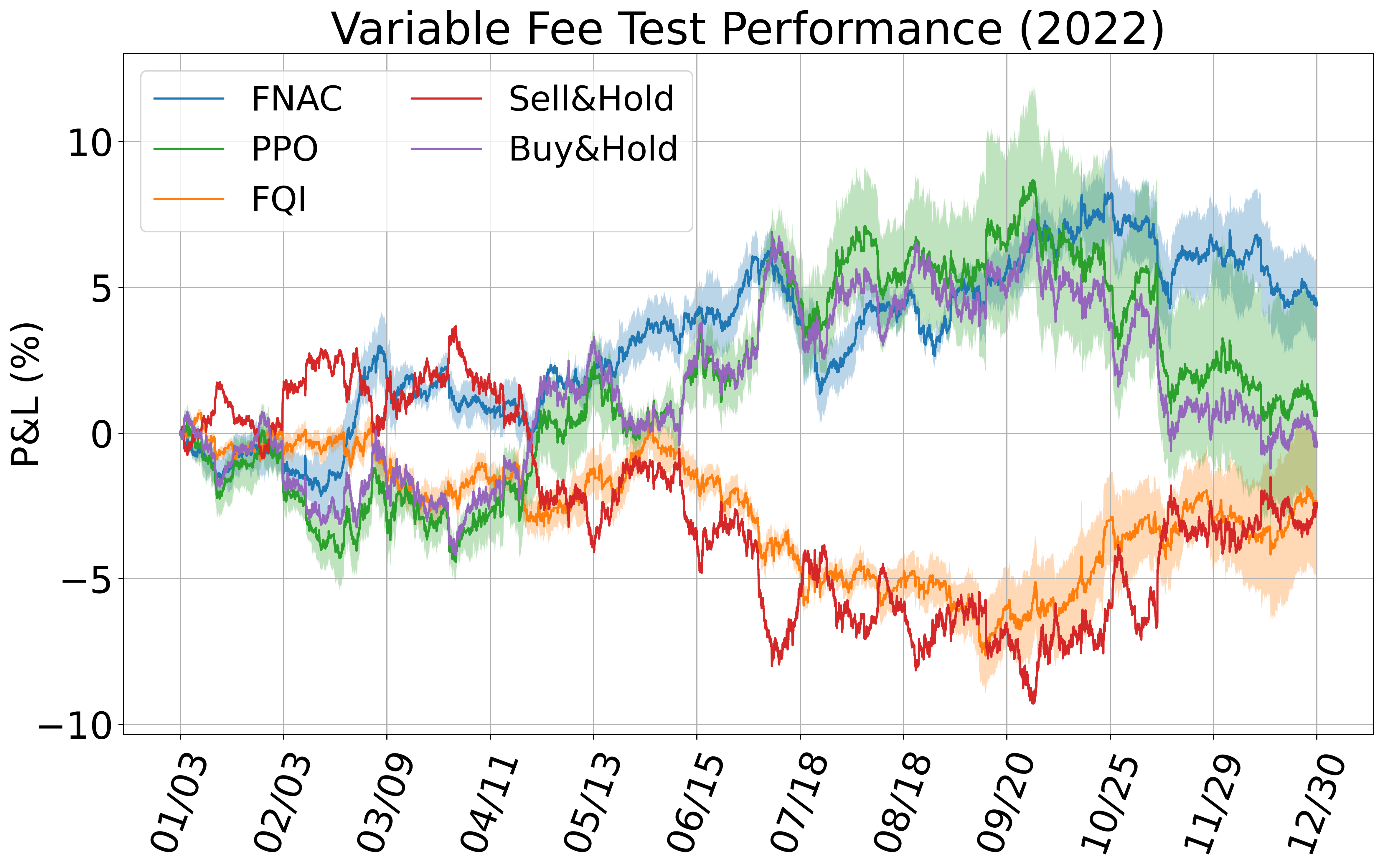}
    \vspace*{1mm}
     \caption{Expected cumulative returns obtained in test by the FNAC model trained in the continuous action space trading setting with variable transaction fees. The performance of FNAC are compared with those obtained by FQI (trained in the discrete action space trading setting), PPO, and the baseline strategies Buy\&Hold and Sell\&Hold. Performance is reported as mean percentages w.r.t. the invested amount (mean $\pm$ 2 stds).}
    \label{fig:2022step_fee_comparision}
    \vspace*{8mm}
\end{figure}
\begin{figure}[t]
    \centering
    \hspace{-1em}
    \includegraphics[width=0.46\textwidth]{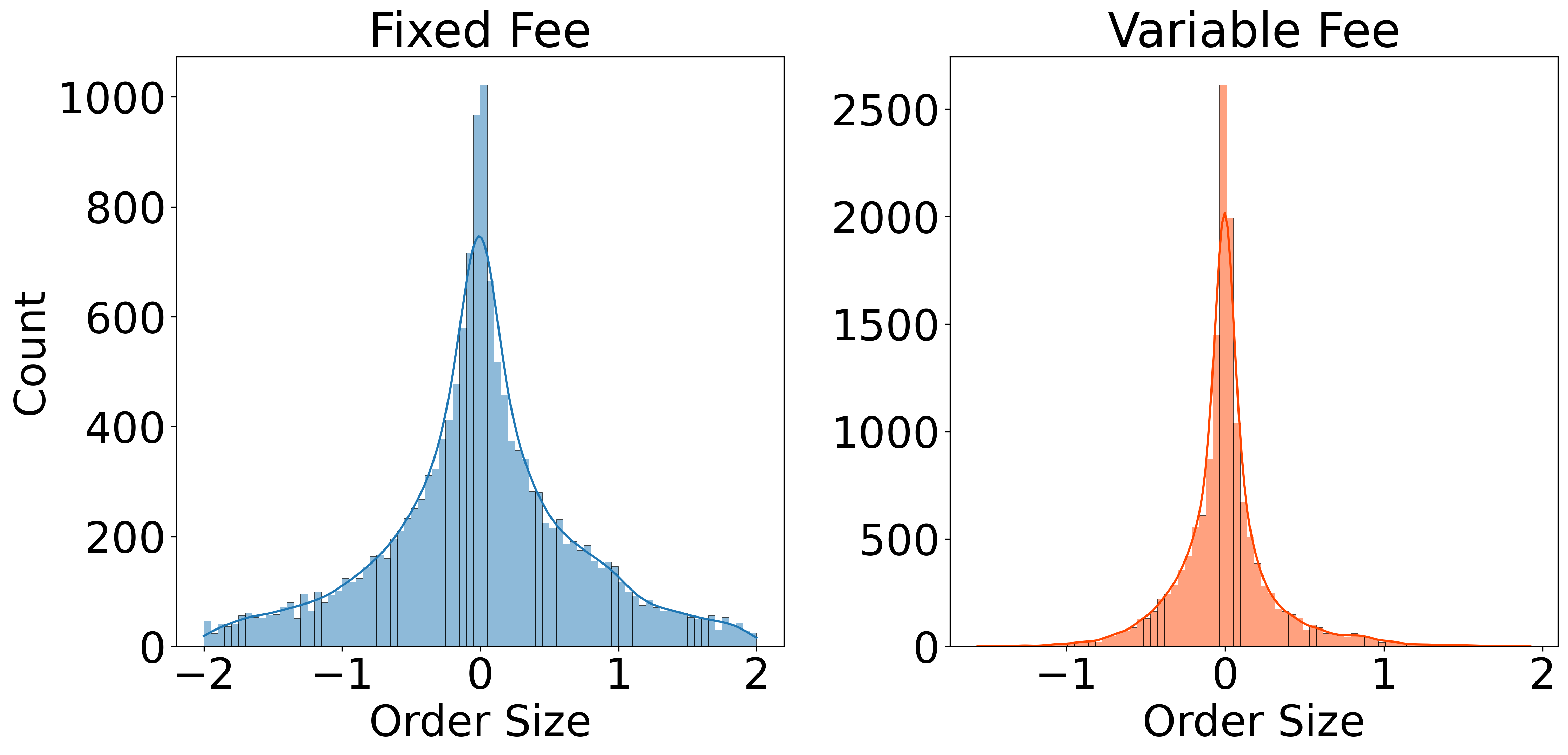}
    \vspace*{1mm}
     \caption{Distribution of the order executed by the FNAC models trained in the continuous action space trading setting with fixed transaction fees (left) and variable transaction fees (right).}
    \label{fig:allocation_changes}
    \vspace*{6mm}
\end{figure}

\begin{figure*}[t]
\centering
\subfloat{
  \includegraphics[width=.75\columnwidth]{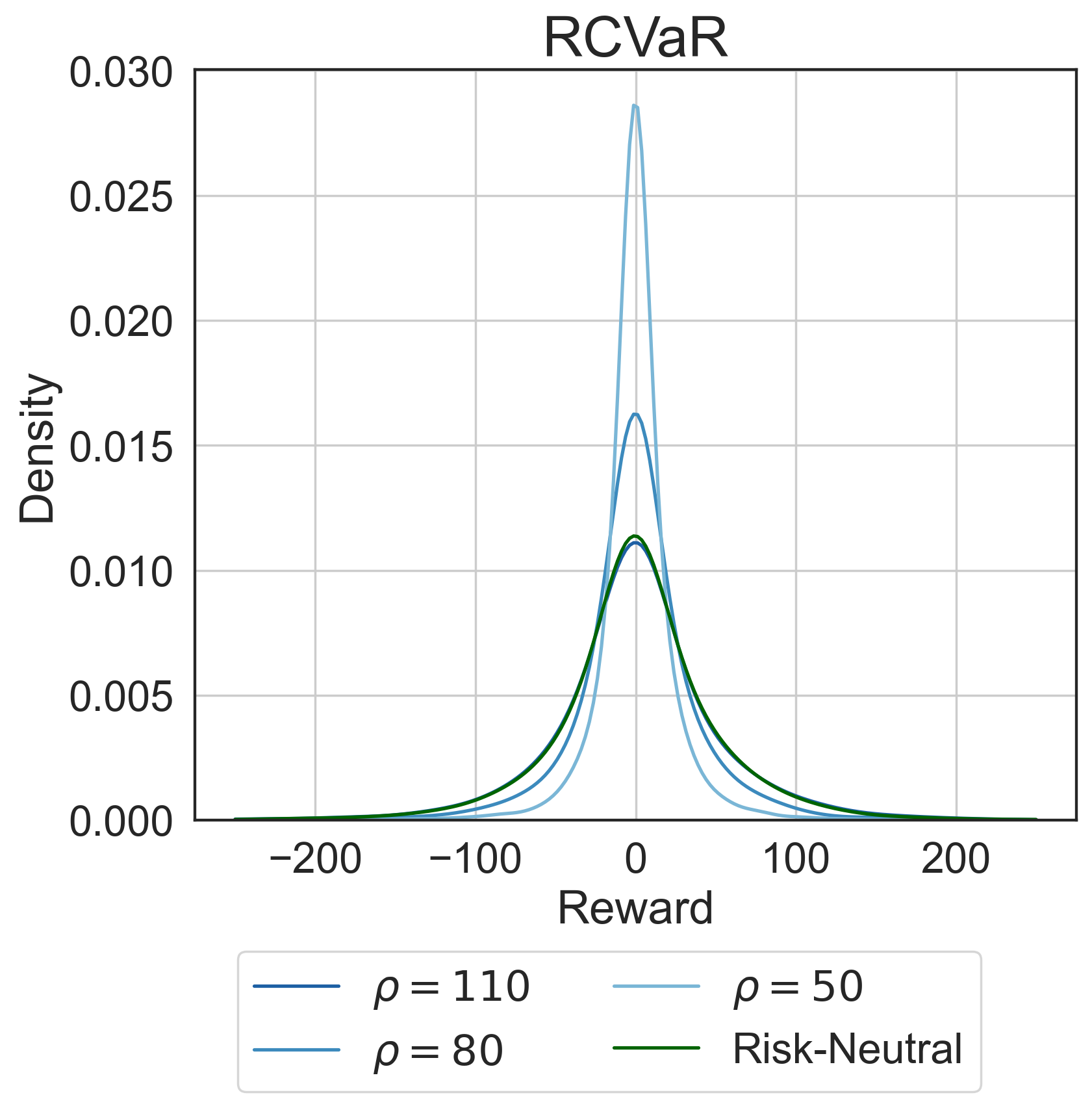}
}
\subfloat{
  \includegraphics[width=.75\columnwidth]{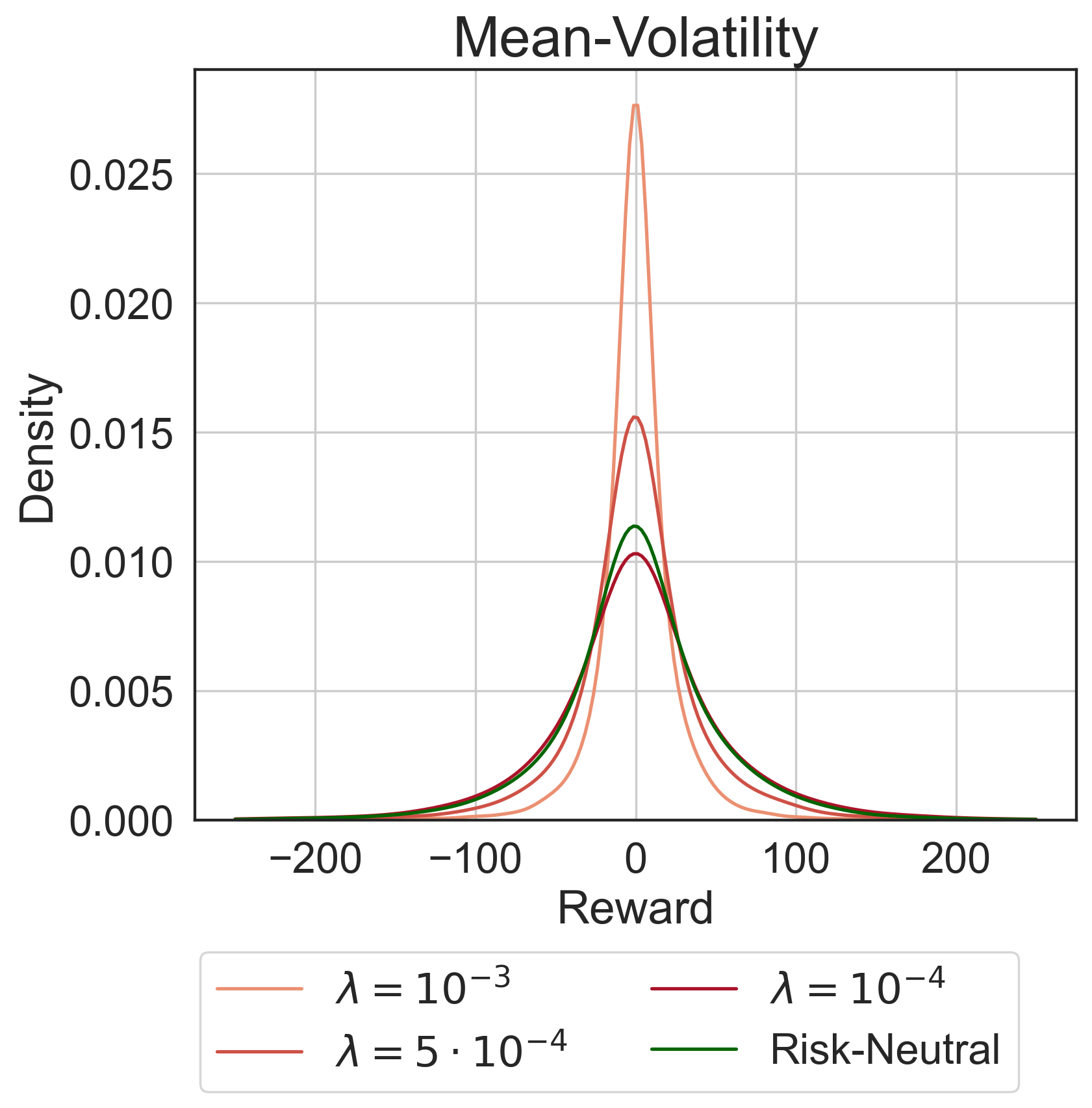}
}
\vspace*{1mm}
\caption{Reward distributions obtained by risk-averse and risk-neutral agents for the test year 2022, fitted by means of a Kernel Density Estimation. Distributions shrinks as risk-aversion increases (i.e. with lower $\rho$ or higher $\lambda$).}
\label{fig:distribution_risk_measures}
\end{figure*}

\subsubsection{Variable Fee Function Setting}
In this section, we deal with the more challenging case of variable transaction costs w.r.t. the order size. In order to consider a realistic cost scenario, we designed the function $g(x)$ in Equation \ref{eq:transaction_cost} as a monotonically increasing step function. In this setting the trader should carefully select the order size in order to balance high fees due to large orders, with a larger return on the investment. With the only exception of the reward function design, the interaction with the environment, the training and validation procedures is straightforward as in the previous continuous setting. As a result, Figure \ref{fig:heatmaps}, 
compares agent allocations in the fixed and the variable fee setting.
The presence of continuous colour stripes shows of how the model has acquired the capacity to incrementally modify the allocations, optimizing the execution of large orders by slicing them in order to avoid large bid-asks thus decreasing the total transaction costs. The same behaviour can be better understood by looking at Figure \ref{fig:allocation_changes}, which shows the difference between two subsequent allocations, demonstrating how the model with the size-dependent function chooses to execute smaller orders.
Figure \ref{fig:2022step_fee_comparision} instead shows the performance of FNAC in this setting, comparing it again with the same aforementioned baselines. The latter algorithm is trained using a discrete action space, which is allowed only to play orders of the maximum size. Thanks to the use of a continuous action space, FNAC manages to achieve a better performance compared to FQI, which is not capable of splitting orders in smaller amounts.
Interestingly, FNAC overperforms PPO also in this case, even if both of them should enjoy, in principle, the advantages of continuous action spaces.

While, at first sight, the performances in the variable scenario may seem worse w.r.t. the fixed fee setting, they cannot actually be compared, as the costs of the former are designed to be higher. Moreover, Moreover, it should be noticed that this scenario considers the possibility of investing larger amounts of money, hence, possibly larger returns in absolute terms. 

\subsubsection{Risk-Averse Setting}
To further explore the possibilities enabled by selecting actions in a continuous action space we study two different types of risk-averse optimization using FNAC.
As previously mentioned, to use RCVaR as a measure of risk we used the transformation of rewards (Equation \ref{eq:rcvar_reward_transformation}). Following \cite{bauerle2011markov} we reverse the usual CVaR perspective, employing $\rho$ as risk level instead of $\alpha$ and, thus, training models with the aforementioned reward transformation. Higher values of the parameter $\rho$ correspond to lower value of risk-aversion.
As shown in Figure \ref{fig:distribution_risk_measures}, lower values of $\rho$ induce a more conservative behaviour, decreasing the variability of the obtained reward distribution.



As the second risk-averse objective is taken into account, we first identified a range of values for the $\lambda$ parameter (around $10^{-3}$) that provided different behaviors. As expected, analyzing the distribution of the reward obtained by the risk-averse model trained with these parameters, it is possible to notice that also in this case higher levels of risk-aversion (i.e. higher values of the $\lambda$ parameter) correspond to a decrease of the variability of the obtained rewards (Figure \ref{fig:distribution_risk_measures}).
\section{Conclusion}\label{sec:conclusions}
Reinforcement Learning offers exciting opportunities for financial applications. 
Successfully learning trading strategies in continuous action spaces is a fundamental step towards the application of this kind of algorithms in real financial tasks. 
On the other hand, the Forex market, thanks to its enormous liquidity and flexibility, is the perfect benchmark to test these automated strategies. 
By adapting the FNAC algorithm to the financial setting, we modelled an agent capable of interacting with the market using continuous actions, exploiting the advantages of employing a parametric actor and a non-parametric critic. 
First, we empirically established that this method can learn effective trading strategy in this more challenging setting. Compared to other state-of-the-art policy based baselines, this method is easier to train and obtains better performance on out-of-sample data. 
Secondly, we exploited the wider range of possibility offered by a continuous action space, validating this approach on other two financial tasks. Through the integration of a size-dependent fee function, we were able to train an agent capable of limiting the order volume in order to better exploit the trade-off between revenues and transaction costs. Furthermore, through the use of risk measures (mean-volatility and RCVaR), the trained agents successfully developed risk-averse behaviors, decreasing the variability of the obtained rewards. Future works could investigate even more complex scenarios, involving multi-dimensional action spaces which allow to trade multiple asset at the same time.
Although positive performances were achieved, the non-stationarity of the data caused performances to deteriorate in the last months of the test year 2022. This can be attributed to the limited ability of the model to generalize, which is partly due to the large variance of the results w.r.t. the policy initialization and the presence of deep differences in the market regimes shown between training and test datasets. Several approaches can be considered to limit the impact of non-stationarity. One of these is the use of multi-expert learning algorithms, through which market drifts can be managed in a better way thanks to the experience of different models. Further possible optimizations result from the computational effort required for the training of the different components of FNAC. This can be done by performing a more in-depth tuning of the hyperparameters used, which is however made difficult by the computational time required. Finally, the differentiable nature of the policy opens up possibilities for investigating transfer learning techniques on currencies other than EUR/USD. This is especially relevant for currencies with lower liquidity, where finding successful strategies can be more challenging.  

\newpage

\bibliography{FNAC}

\end{document}